\documentclass[twocolumn,showpacs,superscriptaddress,prb,aps,floatfix]{revtex4-2}
\usepackage{graphicx}
\usepackage{rotating}
\usepackage{amsmath}
\usepackage[usenames,dvipsnames]{color}
\usepackage{float}
\usepackage{rotating}
\usepackage{booktabs}
\usepackage{multirow}
\usepackage{longtable}
\usepackage[colorlinks,linkcolor=blue,citecolor=blue]{hyperref}
\usepackage[T1]{fontenc}
\usepackage{bm}
\usepackage{upgreek}
\usepackage{comment}

\makeatletter

\newcommand{\Rmnum}[1]{\expandafter\@slowromancap\romannumeral #1@}

\newcommand{\DC}[1]{\textcolor{red}{To Domenico and Cesare:}}
\makeatletter
\hfuzz=\maxdimen
\usepackage{mhchem}

\begin{document}

\title{Ultrahigh Intrinsic Hole Mobilities in $M$N$_2$ ($M$= Mo and W) at Room Temperature}

\author{Zhongjuan Han}
\thanks{$^*$These authors contributed equally to this work.}
\affiliation{Key Laboratory of Advanced Materials and Devices for Post-Moore Chips, Ministry of Education, School of Mathematics and Physics, University of Science and Technology Beijing, Beijing 100083, China}

\author{Rong-Tian Pang}
\thanks{$^*$These authors contributed equally to this work.}
\affiliation{Center for Quantum Physics, Key Laboratory of Advanced Optoelectronic Quantum Architecture and Measurement (MOE), School of Physics, Beijing Institute of Technology, Beijing 100081, China.}

\author{Wu Xiong}
\thanks{$^*$These authors contributed equally to this work.}
\affiliation{Key Laboratory of Advanced Materials and Devices for Post-Moore Chips, Ministry of Education, School of Mathematics and Physics, University of Science and Technology Beijing, Beijing 100083, China}

\author{Zhonghao Xia}
\affiliation{Key Laboratory of Advanced Materials and Devices for Post-Moore Chips, Ministry of Education, School of Mathematics and Physics, University of Science and Technology Beijing, Beijing 100083, China}

\author{Jin-Jian Zhou}
\email{jjzhou@bit.edu.cn}
\affiliation{Center for Quantum Physics, Key Laboratory of Advanced Optoelectronic Quantum Architecture and Measurement (MOE), School of Physics, Beijing Institute of Technology, Beijing 100081, China.}

\author{Jiangang He}
\email{jghe2021@ustb.edu.cn}
\affiliation{Key Laboratory of Advanced Materials and Devices for Post-Moore Chips, Ministry of Education, School of Mathematics and Physics, University of Science and Technology Beijing, Beijing 100083, China}

\date{\today}

	\begin{abstract}
	High-mobility $p$-type semiconductors are essential for advanced electronic devices but remain scarce. Here, using a hierarchical screening framework that combines first-principles calculations with Boltzmann transport theory, we identify $M$N$_2$ ($M$= Mo and W) family as polar semiconductors with exceptionally high intrinsic hole mobilities. In particular, 1H-WN$_2$ exhibits a room-temperature hole mobility exceeding $10^{4}$~$\mathrm{cm^2\,V^{-1}\,s^{-1}}$. This exceptional transport performance arises from the synergistic suppression of polar-optical-phonon and acoustic-phonon scattering, together with a reduced intervalley-scattering phase space induced by spin--valley locking. These effects arise from anomalously small Born effective charges, strong covalent N--N bonds, and orbital hybridization between N-$2p_x$/$2p_y$ and W-$5d_{xy}$/$5d_{x^2-y^2}$ in the N$_2$-dimer-based structure. Our results establish MoN$_2$ and WN$_2$ as a promising class of high-mobility polar semiconductors and introduce a crystal-structure-based strategy for concurrently suppressing multiple electron--phonon scattering channels, thereby revising design principles for high-mobility materials.
	\end{abstract}

	\maketitle
	
	Carrier mobility ($\mu$) quantifies the drift response of electrons ($\mu_{\mathrm{e}}$) or holes ($\mu_{\mathrm{h}}$) to an applied electric field and directly determines key device metrics, including switching speed, electrical conductivity, and power dissipation~\cite{Lundstrom2000}. High-mobility semiconductors thus underpin modern electronic and energy-conversion technologies, from diodes~\cite{sze2021physics} and transistors~\cite{schroder2015semiconductor} to transparent conductors~\cite{hosono2010handbook,10.1063/1.323240}, photovoltaic cells~\cite{nelson2003physics}, and thermoelectric modules~\cite{snyder2008complex}. Identifying materials that combine high $\mu$ with wide band gaps remains a central challenge in electronic-materials research, particularly for $p$-type transport~\cite{10.1021/acsnano.9b07618}.
	
	Within the Drude model~\cite{ashcroft1976solid}, $\mu=e\tau/m^{*}$, where $e$, $\tau$, and $m^{*}$ are the elementary charge, carrier lifetime, and effective mass, respectively. High mobility therefore demands both light carriers and long carrier lifetimes. In defect-free semiconductors at room and elevated temperatures, the intrinsic lifetime is governed primarily by electron--phonon scattering~\cite{peter2010fundamentals,RevModPhys.89.015003,PhysRevB.95.075206,PhysRevB.98.081203,claes2025phonon}. Acoustic deformation-potential (ADP) scattering~\cite{PhysRev.80.72} is often the dominant limitation in nonpolar semiconductors. In polar materials, however, longitudinal-optical (LO) phonons generate macroscopic electric fields that couple to charge carriers through the Fr\"ohlich interaction~\cite{Frohlich01071954}, introducing polar-optical-phonon (POP) scattering as an additional and often dominant $\mu$-limiting channel~\cite{cheng2018limits}.
	
	Therefore, the leading $p$-type semiconductors are predominantly elemental or strongly covalent compounds, including diamond (3800~cm$^2$\,V$^{-1}$\,s$^{-1}$~\cite{doi:10.1126/science.1074374}), black phosphorus ($1000$~cm$^2$\,V$^{-1}$\,s$^{-1}$~\cite{li2014black}), Te ($700$~cm$^2$\,V$^{-1}$\,s$^{-1}$~\cite{wang2018field}), and silicon ($500$~cm$^2$\,V$^{-1}$\,s$^{-1}$~\cite{PhysRevB.97.121201}). In contrast, polar semiconductors generally display much lower room-temperature $\mu_{\mathrm{h}}$, consistent with strong POP scattering. For example, the reported $\mu_{\mathrm{h}}$ values of GaN~\cite{PhysRevLett.123.096602}, Cu$_2$O~\cite{10.1063/1.3026539}, MoSi$_2$N$_4$~\cite{sun2026wafer}, and $\beta$-TeO$_2$~\cite{zavabeti2021high} are 50, 90, 154, and 232~cm$^2$\,V$^{-1}$\,s$^{-1}$, respectively.

    \begin{figure*}[th!]
	\centering
	\includegraphics[width=1\linewidth]{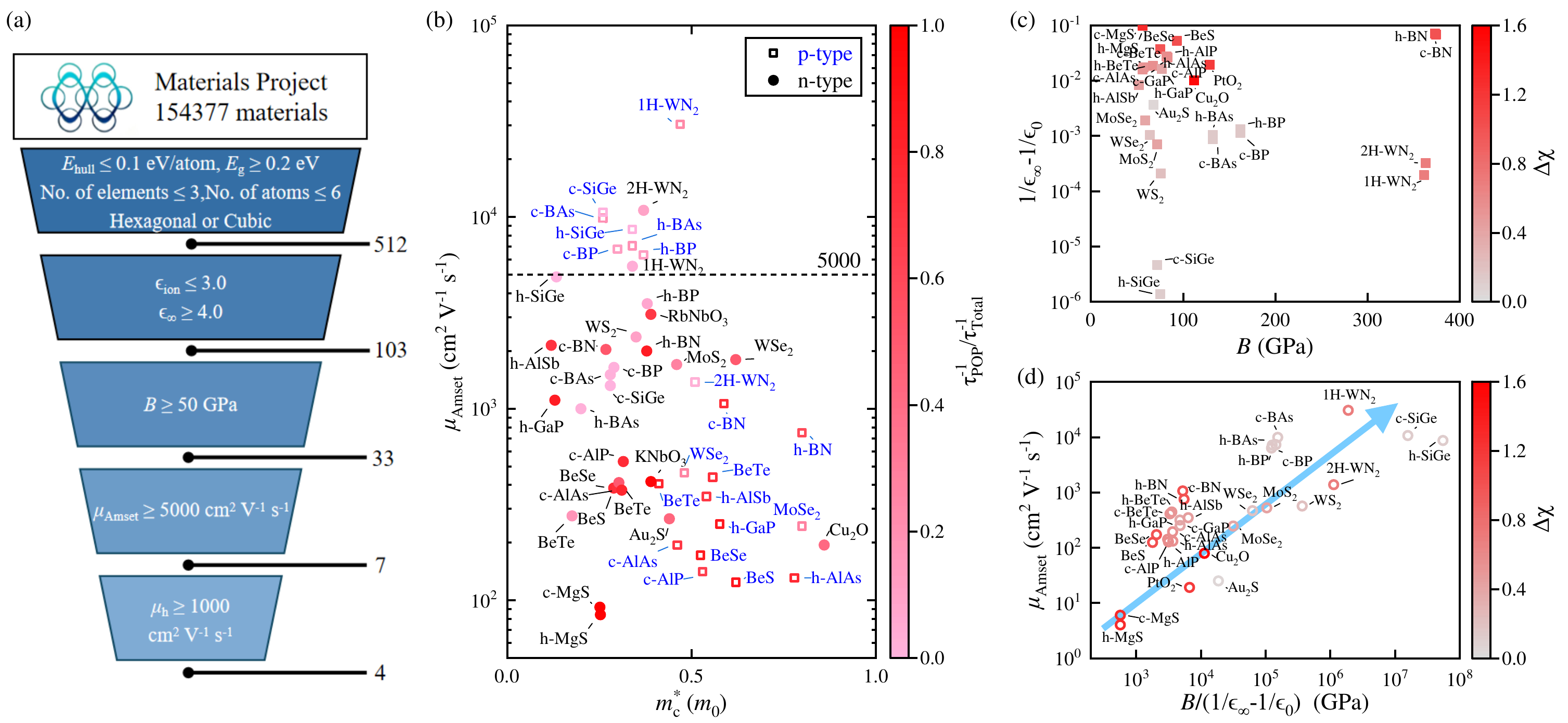}
	\caption{(a) High-throughput screening workflow. (b) AMSET-computed mobility $\mu_{\mathrm{Amset}}$ as a function of the transport effective mass $m^*_\mathrm{c}$, color-coded by the relative contribution of polar-optical-phonon (POP) scattering. (c) Relationship between the dielectric descriptor $(1/\epsilon_\infty - 1/\epsilon_0)$ and the bulk modulus $B$. (d) Correlation between $\mu_{\mathrm{Amset}}$ and the combined descriptor $B/\left(1/\epsilon_\infty - 1/\epsilon_0\right)$, color-coded by the electronegativity difference $\Delta\chi$ between the constituent elements.}
	\label{fig1}
    \end{figure*}

	First-principles calculations have substantially accelerated the search for high-$\mu$ materials, particularly in two-dimensional (2D) semiconductors~\cite{PhysRevB.102.094308,cheng2018limits,10.1021/jacs.9b05923,PhysRevLett.130.087001,ha2024high,10.1021/jacs.6c00402}. Although the reduced dimensionality modifies dielectric screening and the electronic density of states~\cite{PhysRevLett.125.177701}, the central challenge remains the control of electron--phonon scattering. High-throughput first-principles screening combined with electron--phonon calculations and Boltzmann transport theory has identified several promising 2D candidates~\cite{10.1021/jacs.9b05923,cheng2018limits,PhysRevLett.130.087001,ha2024high}. Descriptor-guided screening predicted BSb, ZrI$_2$, HfI$_2$, Sn$_2$H$_2$, and BAs to exhibit $\mu_{\mathrm{h}} \geq 2000$~cm$^2$\,V$^{-1}$\,s$^{-1}$~\cite{PhysRevLett.130.087001}. Complementary high-throughput Boltzmann-transport calculations identified WS$_2$ and hexagonal Sb as promising 2D $p$-type semiconductors with $\mu_{\mathrm{h}}$ approaching 1000~cm$^2$\,V$^{-1}$\,s$^{-1}$~\cite{ha2024high}. These advances establish computational screening as a powerful discovery tool, while underscoring the need for physically transparent descriptors that can preselect materials with weak intrinsic electron--phonon scattering before computationally demanding transport calculations.
	
	In this Letter, we introduce a hierarchical screening strategy that targets polar semiconductors with concurrently weak ADP and POP scattering. From 512 semiconductors, we identify MoN$_2$ and WN$_2$ as exceptionally high intrinsic $\mu_{\mathrm{h}}$ semiconductors at room temperature, with WN$_2$ exhibiting a record-high $\mu_{\mathrm{h}}$ of over 10000~cm$^2$\,V$^{-1}$\,s$^{-1}$. Their ultrahigh $\mu_{\mathrm{h}}$ originates from the simultaneous attenuation of POP and ADP scattering, together with a strongly reduced intervalley-scattering phase space, all enabled by their unusual crystal and electronic structures. Remarkably, the extremely weak POP is unexpected given the substantial electronegativity differences of their constituent elements ($\Delta\chi=0.68$--$0.88$), which conventionally imply polar covalent bonding and strong POP scattering. This unusual mechanism enables polar covalent semiconductors to surpass the $\mu_{\mathrm{h}}$ of conventional covalent, nonpolar semiconductors. Our findings challenge the conventional polarity--mobility tradeoff and establish crystal-structure engineering as a route to control electron--phonon scattering in high-performance semiconductors.

	Within the Fr\"{o}hlich model~\cite{Frohlich01071954,PhysRevLett.115.176401}, the electron--phonon coupling matrix element for scattering by LO phonons scales as $(1/\epsilon_\infty-1/\epsilon_0)$, where $\epsilon_\infty$ and $\epsilon_0$ are the high-frequency and static dielectric constants, respectively. Because the static dielectric response contains electronic and ionic contributions, $\epsilon_0=\epsilon_\infty+\epsilon_{\mathrm{ion}}$, the POP interaction is weakened by a small ionic dielectric contribution, $\epsilon_{\mathrm{ion}}$, and a sufficiently large electronic dielectric constant, $\epsilon_\infty$. In contrast, for ADP scattering~\cite{PhysRev.80.72}, carrier mobility increases with the relevant elastic constant, $c_{ii}$. For high-throughput screening, this directional elastic response can be conveniently approximated by the bulk modulus, $B$.
	
	Guided by these physical considerations, we developed a hierarchical screening workflow to identify high-$\mu$ semiconductors through the simultaneous suppression of POP and ADP scattering. As illustrated in Fig.~\ref{fig1}(a), $\epsilon_\infty$, $\epsilon_{\mathrm{ion}}$, and $B$ were used as screening descriptors. We began with compounds in the Materials Project~\cite{10.1063/1.4812323} and retained thermodynamically stable materials ($E_\mathrm{hull}$ = 0.0~eV/atom) and metastable materials ($E_\mathrm{hull} \leq 0.1$~eV/atom) with a PBE~\cite{PBE} band gap $E_{\mathrm{g}}>0.2$~eV. To enable efficient high-throughput calculations, we further restricted the search to hexagonal and cubic compounds with no more than six atoms per unit cell and no more than three distinct chemical elements. This initial selection yielded 512 candidates.
	
	We next imposed $\epsilon_{\mathrm{ion}}\leq 3$ and $\epsilon_\infty\geq 4$, reducing the candidate set to 103 materials. Requiring $B\geq 50$~GPa further narrowed the set to 33 compounds. We then evaluated the room-temperature $\mu$ of these 33 semiconductors using the AMSET package~\cite{amset}, which includes both POP and ADP scattering; the resulting values are denoted by $\mu_{\mathrm{Amset}}$. For candidates with $\mu_{\mathrm{Amset}}\geq 5000$~cm$^2$\,V$^{-1}$\,s$^{-1}$ at room temperature, we performed higher-accuracy first-principles transport calculations by solving the electron Boltzmann transport equation with the full electron--phonon interaction included. These calculations include spin--orbit coupling (SOC) and quadrupole corrections~\cite{PhysRevB.102.125203,PhysRevB.102.094308} and were carried out using the Perturbo code~\cite{Perturbo}. Computational details are provided in \textcolor{red}{Note S1} of the Supplemental Material~\cite{Sl} and references therein~\cite{amset,vasp1,vasp2,PBE,BoltzTraP2,LOBSTER,irvsp,QE,ONCV,mbj,wannier90,Perturbo,phonopy,CSLD,alamode,alamode2,scph1, scph2,scph3,HAN2022108179,ShengBTE_2014,guo2024sampling,simoncelli2019unified}.

	\begin{figure*}[th!]
		\centering
		\includegraphics[width=1\linewidth]{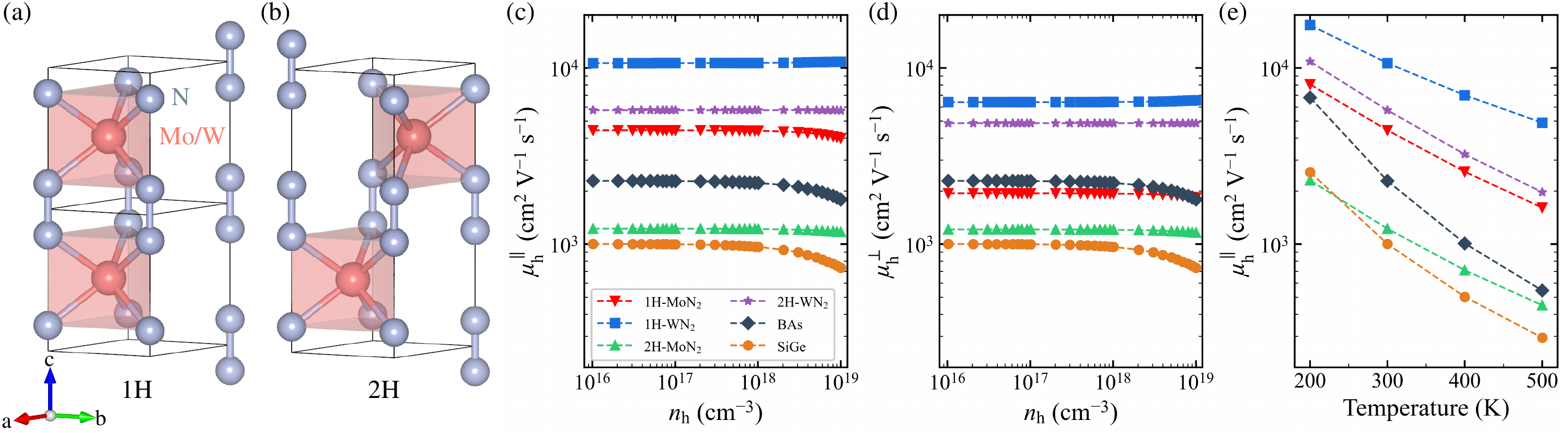}
		\caption{(a,b) Crystal structures of 2H- and 1H-$M$N$_2$, respectively. (c,d) Out-of-plane ($\mu_{\mathrm{h}}^{\parallel}$) and in-plane ($\mu_{\mathrm{h}}^{\perp}$) hole mobilities, calculated with the full electron--phonon coupling, as functions of the hole concentration $n_{\mathrm{h}}$. (e) Temperature dependence of the out-of-plane hole mobility $\mu_{\mathrm{h}}^{\parallel}$ for the materials investigated in this work.} 
		\label{fig2} 
	\end{figure*}

	Figure~\ref{fig1}(b) maps the AMSET calculated mobility ($\mu_{\mathrm{Amset}}$) against the transport effective mass ($m^*_\mathrm{c}$) for both $n$- and $p$-type doping in the 33 screened semiconductors. For hexagonal systems, only the in-plane ($\mu^{\perp}$) component of mobility is shown. As expected from the Drude model~\cite{ashcroft1976solid}, $\mu_{\mathrm{Amset}}$ generally decreases with increasing $m^*_\mathrm{c}$. At comparable $m^*_\mathrm{c}$, materials with a smaller ratio of POP to the total scattering rate ($\tau^{-1}_{\mathrm{POP}}/\tau^{-1}_{\mathrm{total}}$) exhibit higher $\mu$, confirming that POP scattering is the dominant $\mu$-limiting mechanism in polar semiconductors.
	
	Among the 33 candidates, cubic (c) and hexagonal (h) SiGe, BP, BAs, and 1H- and 2H-WN$_2$ exhibit $\mu_{\mathrm{Amset}}$ $\geq$ 5000~cm$^2$\,V$^{-1}$\,s$^{-1}$. The high room-temperature $\mu$ of BP and BAs are well established and originate from their weak POP scattering~\cite{PhysRevB.98.081203}, consistent with their predominantly covalent bonding and small electronegativity differences ($\Delta\chi=0.15$ and 0.14 for BP and BAs, respectively). Remarkably, 1H- and 2H-WN$_2$ yield even higher $\mu_{\mathrm{Amset}}$ values than BP and BAs, despite the larger $m^*_\mathrm{c}$ of 1H-WN$_2$ ($0.44~m_0$) relative to BAs ($0.30~m_0$). This enhancement originates from the exceptionally weak POP scattering in WN$_2$. Such behavior is counterintuitive for transition-metal nitrides, for which the high electronegativity of N would normally be expected to produce appreciable ionicity and strong POP scattering. The anomalously weak POP interaction is therefore a central ingredient of the ultrahigh $\mu_{\mathrm{h}}$ in 1H- and 2H-WN$_2$.
	
	To expose the underlying materials trend, Fig.~\ref{fig1}(c) plots $(1/\epsilon_\infty-1/\epsilon_0)$ against $B$ for the 33 screened compounds. Both 1H- and 2H-WN$_2$ occupy an unusual regime combining a large bulk modulus with an ultralow $(1/\epsilon_\infty-1/\epsilon_0)$, despite their relatively large $\Delta\chi$. In SiGe, the small $\Delta\chi$ (0.11) leads to an ultralow $(1/\epsilon_\infty-1/\epsilon_0)$ and, hence, negligible POP scattering. Its more moderate $B$, however, enhances ADP scattering and results in a lower $\mu_{\mathrm{Amset}}$ than in 1H-WN$_2$. Conversely, BN possesses an exceptionally large $B$ but a much larger $(1/\epsilon_\infty-1/\epsilon_0)$ due to its large $\Delta\chi$ (1.0). The resulting strong POP scattering limits $\mu_{\mathrm{Amset}}$ to approximately $1000$~cm$^2$\,V$^{-1}$\,s$^{-1}$ at room temperature.

	Figure~\ref{fig1}(d) further shows that $\mu_{\mathrm{Amset}}$ generally increases with $B/(1/\epsilon_\infty-1/\epsilon_0)$, reflecting the concomitant suppression of ADP and POP scattering. This relationship, however, becomes nonmonotonic in the high-$\mu_{\mathrm{Amset}}$ regime: although SiGe possesses the largest value of $B/(1/\epsilon_\infty-1/\epsilon_0)$, WN$_2$ exhibits the highest $\mu_{\mathrm{Amset}}$. This deviation arises because the POP-scattering contribution saturates once $(1/\epsilon_\infty-1/\epsilon_0)$ becomes sufficiently small. In particular, reducing this quantity below approximately $10^{-5}$ yields little further enhancement of $\mu_{\mathrm{Amset}}$, such that ADP scattering and effective mass become the dominant factors governing mobility. We note that the effect of the deformation potential in the ADP model~\cite{PhysRev.80.72} is not included by $B/(1/\epsilon_\infty-1/\epsilon_0)$. Therefore, although $B/(1/\epsilon_\infty-1/\epsilon_0)$ is not a complete descriptor, it provides a simple and useful first-pass screening metric for polar semiconductors. Quantitative mobility predictions nevertheless require consideration of the full carrier-scattering landscape.

	WN$_2$ was first predicted by Wang \textit{et al.}~\cite{PhysRevB.79.132109} in the $P\bar{6}m2$ structure and is collected in the Materials Project~\cite{10.1063/1.4812323} as a metastable phase with a convex-hull distance of 86~meV/atom. This phase is nearly degenerate with $P6_3/mmc$ phase, with an energy difference of only 0.3~meV/atom~\cite{PhysRevB.79.132109,SOTO20121}. As illustrated in Fig.~\ref{fig2}(a) and (b), these polymorphs differ solely in the stacking sequence of edge-sharing WN$_6$ triangular-prism sheets: $P\bar{6}m2$ and $P6_3/mmc$ correspond to $AA$ and $AB$ stacking and are thus denoted 1H and 2H, respectively. MoN$_2$ was synthesized under high pressure~\cite{10.1021/jacs.5b01446}, and its phase was subsequently identified as $P6_3/mmc$~\cite{10.1021/acs.jpcc.6b00665}. This 2H-MoN$_2$ polymorph is isostructural with 2H-WN$_2$ and is thermodynamically stable in OQMD~\cite{kirklin2015open}. Because neither 1H- nor 2H-MoN$_2$ is currently available in the Materials Project~\cite{10.1063/1.4812323}, MoN$_2$ was not identified by our initial high-throughput screening. We therefore performed state-of-the-art electron Boltzmann transport calculations to determine the phonon-limited $\mu_{\mathrm{h}}$ of 1H- and 2H-MoN$_2$ and WN$_2$, using BP, BAs, and SiGe as reference materials.

	\begin{figure}[th!]
		\centering
		\includegraphics[width=1\linewidth]{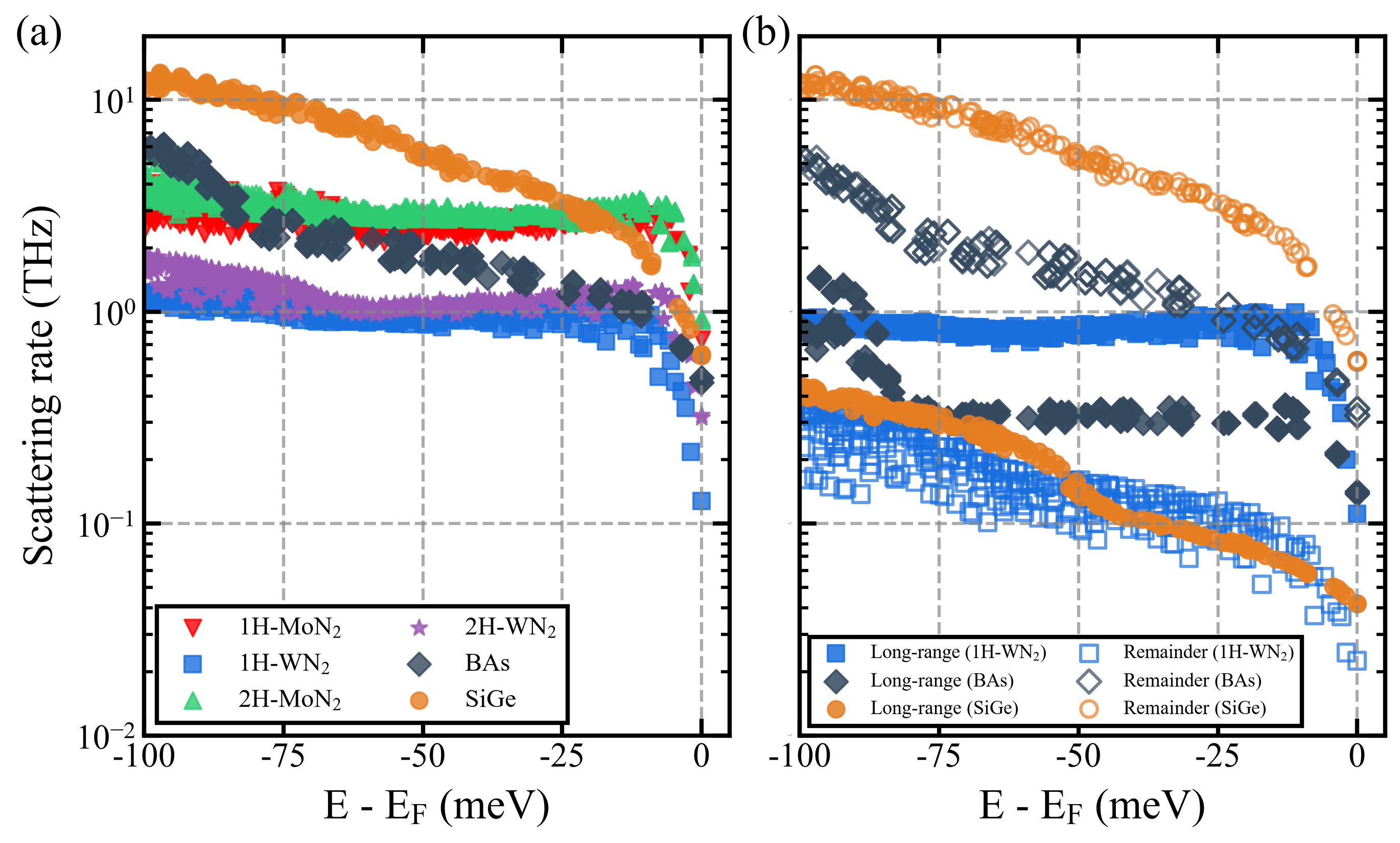}
		\caption{(a) Room-temperature total electron--phonon scattering rates in 1H/2H-$M$N$_2$, BAs, and SiGe, including the quadrupolar correction. (b) Decomposition of the electron--phonon scattering rates into long-range and remainder contributions for 1H-WN$_2$, BAs, and SiGe, including the quadrupolar correction.}
		\label{fig3}
	\end{figure}

    The calculated $\mu_{\mathrm{h}}$ by combining first-principles electron--phonon scattering with Boltzmann transport theory as implemented in Perturbo~\cite{Perturbo}, are shown in Fig.~\ref{fig2}(c) and (d). At room temperature, both the in-plane ($\mu_{\mathrm{h}}^\perp$) and out-of-plane ($\mu_{\mathrm{h}}^{\parallel}$) mobilities of 1H- and 2H-WN$_2$ exceed 4000~cm$^2$\,V$^{-1}$\,s$^{-1}$, surpassing those of the more covalent benchmark material BAs. Among the compounds considered, 1H-WN$_2$ exhibits the highest room-temperature $\mu_{\mathrm{h}}$, reaching $\mu_{\mathrm{h}}^\perp$ 6404~cm$^2$\,V$^{-1}$\,s$^{-1}$ and $\mu_{\mathrm{h}}^{\parallel}$ 10659~cm$^2$\,V$^{-1}$\,s$^{-1}$. It is followed by 2H-WN$_2$, with $\mu_{\mathrm{h}}^\perp$ 4864~cm$^2$\,V$^{-1}$\,s$^{-1}$ and $\mu_{\mathrm{h}}^{\parallel}$ = 5763~cm$^2$\,V$^{-1}$\,s$^{-1}$. Although 2H-MoN$_2$ has the lowest $\mu_{\mathrm{h}}$ in the $M$N$_2$ family, its room-temperature $\mu_{\mathrm{h}}$ remains above 1000~cm$^2$\,V$^{-1}$\,s$^{-1}$. The lower $\mu_{\mathrm{h}}$ of 1H- and 2H-MoN$_2$ relative to their WN$_2$ counterparts originates from the weaker SOC of Mo, which produces a smaller band splitting (see \textcolor{red}{Note~S2} of the Supplemental Material~\cite{Sl}) and thus weaker intervalley-scattering suppression. In addition, the systematically lower $\mu_{\mathrm{h}}$ of 2H-$M$N$_2$ compared with their 1H counterparts arises from the enhanced band degeneracy at the H point and the slightly larger $m_{\mathrm{c}}^{*}$ of the 2H phases, see \textcolor{red}{Note~S3} of the Supplemental Material~\cite{Sl}. Over the hole-concentration range of $10^{16}$--$10^{18}$~cm$^{-3}$, the mobilities of all compounds remain nearly constant. A modest reduction emerges between $10^{18}$ and $10^{19}$~cm$^{-3}$, where the increased density of accessible electronic states away from the Fermi level enhances electron--phonon scattering. As shown in Fig.~\ref{fig2}(e), $\mu_{\mathrm{h}}$ displays the pronounced temperature dependence characteristic of phonon-limited transport. For example, $\mu_{\mathrm{h}}^{\parallel}$ of 1H-WN$_2$ decreases from 17533~cm$^2$\,V$^{-1}$\,s$^{-1}$ at 200~K to 4891~cm$^2$\,V$^{-1}$\,s$^{-1}$ at 500~K. Remarkably, the $\mu_{\mathrm{h}}$ remains exceptionally high even at 500~K, demonstrating the robustness of phonon-limited hole transport in 1H-WN$_2$ substantially above room temperature.

    The corresponding electron--phonon scattering rates including long-range quadrupolar electron--phonon interactions~\cite{PhysRevB.102.125203,PhysRevB.102.094308} are shown in Fig.~\ref{fig3}(a), whereas the ones without quadrupolar correction is shown in \textcolor{red}{Note S4} of the Supplemental Material~\cite{Sl}. Near the Fermi level, 1H-WN$_2$ exhibits the smallest total scattering rate, approximately $0.1$~THz, nearly five times lower than that of BAs. The scattering rate then increases in the sequence 2H-WN$_2$, BAs, SiGe, 1H-MoN$_2$, and 2H-MoN$_2$. This ordering closely mirrors the inverse trend in $\mu_{\mathrm{h}}$, with SiGe as the sole exception, highlighting electron--phonon scattering as the primary determinant of mobility. The total scattering rate can be decomposed into a long-range component, dominated by POP scattering, and a remainder component, arising primarily from ADP scattering~\cite{PhysRevB.94.201201}. Figure~\ref{fig3}(b) compares these contributions for 1H-WN$_2$, BAs, and SiGe. Relative to BAs, 1H-WN$_2$ displays stronger long-range scattering but a markedly weaker remainder contribution. Its stronger long-range scattering, which is different from the descriptor-based trend, originates from the quadrupolar correction in 1H-WN$_2$. In contrast, the remainder contribution dominates in SiGe, consistent with its small $(1/\epsilon_\infty-1/\epsilon_0)$ and moderate $B$, as anticipated from the preceding analysis. To further clarify these trends, we calculate $\mu_{\mathrm{h}}$ for $M$N$_2$ and BAs by retaining only the long-range interaction. As listed in \textcolor{red}{Note~S5} of the Supplemental Material~\cite{Sl}, neglecting the quadrupolar correction yields long-range-limited mobilities in $M$N$_2$ that are far larger than the corresponding total mobilities, demonstrating that the remainder contribution dominates the overall scattering. Once the quadrupolar correction is included, however, the long-range-limited mobilities decrease sharply and become comparable to the total $\mu_{\mathrm{h}}$. In BAs, by contrast, the mobility limited by long-range scattering remains substantially larger than the total mobility both with and without the quadrupolar correction, confirming the dominant role of remainder scattering channels.

    The exceptionally weak POP and ADP scattering in 1H- and 2H-$M$N$_2$ originates from their distinctive crystal and electronic structures. Their small $\epsilon_{\mathrm{ion}}$ is associated with anomalously reduced Born effective charges of $M$, far below those anticipated from either the nominal ionic charges ($M^{4+}$, [N$_2$]$^{4-}$) or the Bader charges; see \textcolor{red}{Note S6} of the Supplemental Material~\cite{Sl}. This reduction is further reinforced by the ultrahigh optical-phonon frequencies~\cite{BornHuang1954}, which arise from the strong N--N covalent bond within the N$_2$ dimer; see \textcolor{red}{Note S7} of the Supplemental Material~\cite{Sl}. As illustrated in Fig.~\ref{fig2}(a) and (b), each $M$ atom is coordinated by six N atoms in a trigonal-prismatic environment, analogous to that of MoS$_2$. In marked contrast to MoS$_2$, however, where neighboring layers are coupled predominantly through van der Waals interactions, both 1H- and 2H-$M$N$_2$ feature strong covalent N--N bonds that bridge adjacent layers. The crystal orbital bond index (COBI)~\cite{10.1021/acs.jpcc.1c00718} of the N--N bond reaches 1.0, confirming its strong covalent character. This bonding markedly reduces the interlayer spacing; see \textcolor{red}{Note S3} of the Supplemental Material~\cite{Sl}, and provides the structural basis for the exceptionally high incompressibility~\cite{PhysRevB.79.132109}. Similar N$_2$-dimer-containing structures have also been identified in other transition-metal dinitrides~\cite{10.1021/acs.inorgchem.5c01118}. To isolate the role of this N$_2$-based structural motif, we further calculate $\epsilon_{\mathrm{ion}}$, $(1/\epsilon_\infty-1/\epsilon_0)$, and $B$ for other WN$_2$ polymorphs without N$_2$ dimers, as collected in the Materials Project~\cite{10.1063/1.4812323}. As summarized in \textcolor{red}{Note S8} of the Supplemental Material~\cite{Sl}, these structures exhibit substantially lower $B$ and much larger $\epsilon_{\mathrm{ion}}$, while retaining comparable $\epsilon_\infty$. Consequently, they display much smaller $B/(1/\epsilon_\infty-1/\epsilon_0)$ and correspondingly lower $\mu_{\mathrm{Amset}}$.

	Our electronic-structure calculations show that all four $M$N$_2$ compounds are indirect-gap semiconductors with the valence-band maximum (VBM) at the H point of the first Brillouin zone, where pronounced hybridization between $M$-$d$ and N-2$p$ states occurs. Calculations using the mBJ~\cite{mbj} exchange potential including SOC yield band gaps of 0.97 and 0.95~eV for 1H- and 2H-MoN$_2$, respectively, and 1.15 and 1.19~eV for 1H- and 2H-WN$_2$, respectively, see \textcolor{red}{Note S3} of the Supplemental Material~\cite{Sl}. These gaps are comparable to that of Si (1.17~eV at 0~K)~\cite{PhysRevLett.55.1418}. Because the four compounds share closely related electronic structures, the following discussion focuses on 1H-WN$_2$.
	
\begin{figure}[th!]
		\centering
		\includegraphics[width=1\linewidth]{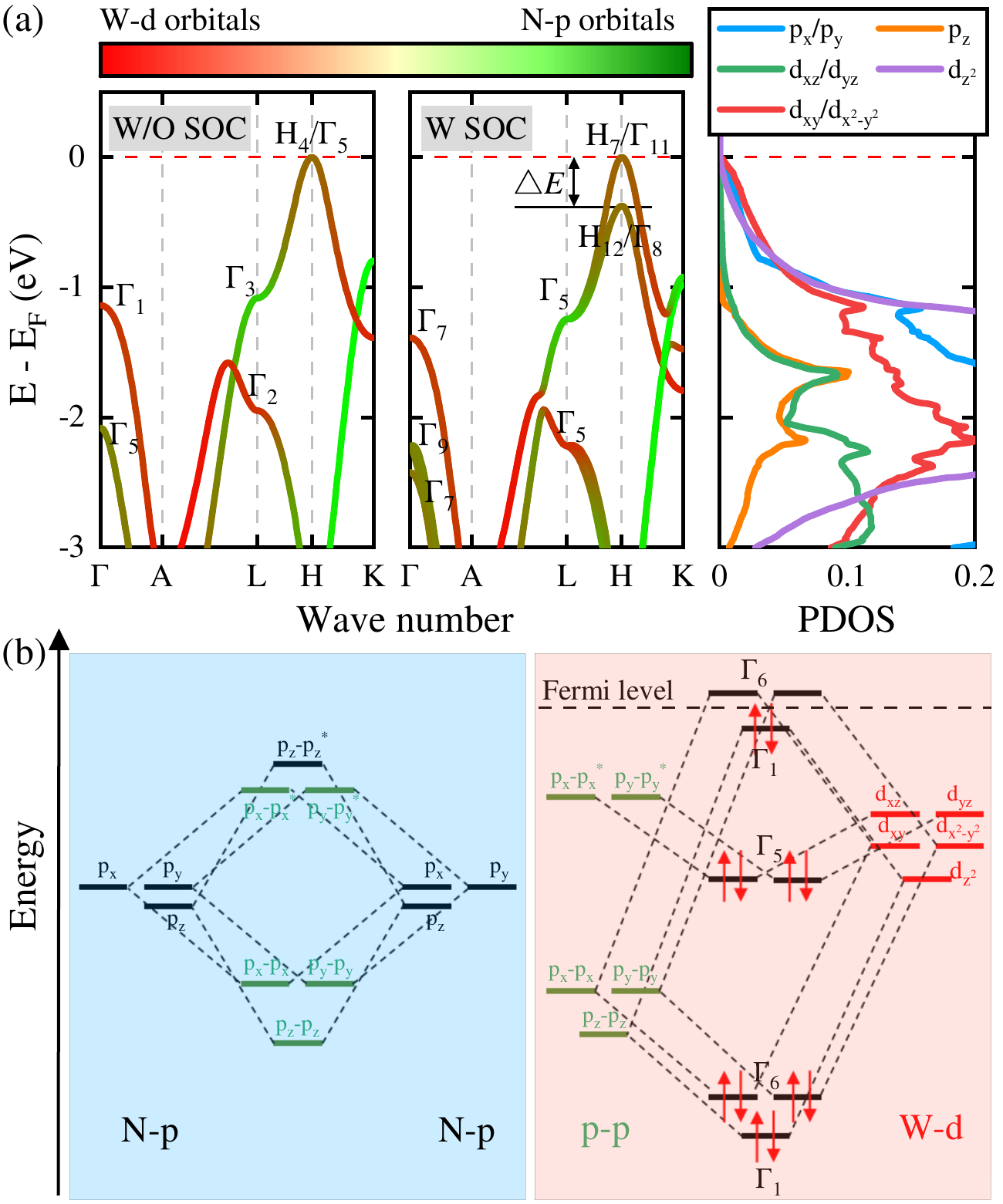}
		\caption{(a) Orbital-resolved electronic band structures of 1H-WN$_2$ calculated without and with spin--orbit coupling (SOC), labeled by the irreducible representations (irreps) at the relevant high-symmetry $k$ points, together with the orbital-projected density of states (PDOS) obtained without SOC. (b) Molecular-orbital diagram of 1H-WN$_2$ at the $\Gamma$ point.}
		\label{fig4}
	\end{figure}

	Figure~\ref{fig4}(a) shows the valence-band structure of 1H-WN$_2$ without and with SOC; the complete band structures of all four compounds are provided in \textcolor{red}{Note S2} of the Supplemental Material~\cite{Sl}. SOC induces a substantial splitting of the valence-band states at H. This splitting suppresses electron--phonon scattering through spin--valley locking~\cite{ha2025ultrahigh}, enhancing $\mu_{\mathrm{h}}^{\parallel}$ by a factor of 10, from 976 to 10659~cm$^2$\,V$^{-1}$\,s$^{-1}$. As shown in \textcolor{red}{Note S9} of the Supplemental Material~\cite{Sl}, states near the VBM are dominated by N-2$p_x$/2$p_y$ and W-5$d_{xy}$/5$d_{x^2-y^2}$ orbitals. At deeper energies, from approximately $-1$ to $-3$~eV, the W-5$d_{z^2}$ contribution becomes dominant over the W-5$d_{xy}$/5$d_{x^2-y^2}$ contributions, as is particularly evident in the projected density of states in Fig.~\ref{fig4}(a).
	
	According to the symmetry-matching principle of molecular-orbital theory~\cite{PhysRev.32.186,hund1927deutung}, atomic orbitals can hybridize to form molecular orbitals only when they share compatible band representations (BRs)~\cite{PhysRevE.96.023310,TQC,xiong2025forbidden}. Using the Irvsp code~\cite{irvsp}, we find that the VBM of 1H-WN$_2$ without SOC transforms as the H$_4$ irreducible representation (irrep) of the space group. This irrep is induced exclusively by the BR $\mathrm{H_3\oplus H_4\oplus H_5\oplus H_6}$ associated with the N-2$p_x$/2$p_y$ orbitals and the BR $\mathrm{H_2\oplus H_4}$ associated with the W-5$d_{xy}$/5$d_{x^2-y^2}$ orbitals, see \textcolor{red}{Note S10} of the Supplemental Material~\cite{Sl}. Consequently, the VBM derives solely from these orbital manifolds, and the relevant hybridization is restricted to their symmetry-allowed interactions.

	We further constructed a molecular-orbital diagram for 1H-WN$_2$ in the absence of SOC, see Fig.~\ref{fig4}(b) and \textcolor{red}{Note S11} in the Supplemental Material~\cite{Sl}. This analysis shows that the N$_2$ dimer acts as an integrated molecular unit: the two N-$2p_z$ orbitals first hybridize to form molecular orbitals, which subsequently interact with the W-$5d$ orbitals. At the $\Gamma$ point, the highest valence band has a weak net antibonding character, arising from antibonding $d_{z^2}$--$p_z$ interactions partially compensated by bonding $p_z$--$p_z$ contributions. By contrast, the VBM at H comprises a superposition of strong W--N antibonding states ($d_{xy}$--$p_x$, $d_{xy}$--$p_y$, $d_{x^2-y^2}$--$p_x$, and $d_{x^2-y^2}$--$p_y$) and N--N antibonding states ($p_x$--$p_x$ and $p_y$--$p_y$). Quantitatively, although the W--N antibonding interaction is stronger at $\Gamma$ ($-\mathrm{ICOHP}=0.59$) than at H ($-\mathrm{ICOHP}=0.50$), the N--N interaction is bonding at $\Gamma$ ($-\mathrm{ICOHP}=-0.43$) but antibonding at H ($-\mathrm{ICOHP}=0.24$). Consequently, the total antibonding interaction at H ($-\mathrm{ICOHP}=0.74$) is 4.6 times stronger than that at $\Gamma$ ($-\mathrm{ICOHP}=0.16$). This pronounced contrast places the VBM at H, while pushing the corresponding $\Gamma$-point state approximately 1.1~eV lower in energy.
	
	The N--N bonds thus play two pivotal roles in the 1H- and 2H-$M$N$_2$ compounds. First, their strong covalency enhances $B$ and raises the optical-phonon frequencies. The resulting suppression of ADP and POP scattering provides a robust structural basis for the high $\mu_{\mathrm{h}}$. Second, the contrasting bonding characters of the N--N interactions at the $\Gamma$ and H shift the VBM from $\Gamma$ to H, thereby establishing the electronic foundation for spin splitting and spin--valley locking in these compounds.
	
	The SOC-induced band splitting is dictated by the little-group symmetries at the relevant high-symmetry points. The point groups at $\Gamma$, L, and H are D$_{\mathrm{3h}}$, C$_{\mathrm{2v}}$, and C$_{\mathrm{3h}}$, respectively. The character table of the double group D$^\mathrm{D}_{\mathrm{3h}}$ is provided in \textcolor{red}{Note S12} of the Supplemental Material~\cite{Sl}. Taking the $\Gamma$ point as an example, the spin character of D$_{\mathrm{3h}}$ associated with the $D^{1/2}$ representation is $\{2,-2,0,1,-1,\sqrt{3},-\sqrt{3},0,0\}$. Taking the direct products of this spin representation with the orbital irreps $\Gamma_1$ and $\Gamma_5$ gives
	\begin{align}
		\Gamma_1 \otimes D^{1/2} &= \Gamma_7, \\
		\Gamma_5 \otimes D^{1/2} &= \Gamma_7 \oplus \Gamma_9.
	\end{align}
	Thus, upon including SOC, the nondegenerate $\Gamma_1$ band transforms into the doubly degenerate $\Gamma_7$ band and remains unsplit, whereas the doubly degenerate $\Gamma_5$ band separates into the $\Gamma_7$ and $\Gamma_9$ bands. The same symmetry analysis applies to the L and H points. Crucially, the double group C$^\mathrm{D}_{\mathrm{2v}}$ contains only one additional two-dimensional irrep, $\Gamma_5$, relative to C$_{\mathrm{2v}}$, whereas all irreps of C$^\mathrm{D}_{\mathrm{3h}}$ are one-dimensional. Consequently, bands at $k$ points with C$_{\mathrm{2v}}$ symmetry, such as L, remain unsplit under SOC, whereas those at $k$ points with C$_{\mathrm{3h}}$ symmetry, such as H, split. These symmetry-derived conclusions are fully consistent with our calculated band structures.
	
	Finally, because efficient heat removal is essential for electronic applications, we calculated the lattice thermal conductivities $\kappa_{\mathrm{L}}$ of the four $M$N$_2$ compounds using the unified theory of thermal transport~\cite{simoncelli2019unified}; computational details are provided in \textcolor{red}{Note S1} of the Supplemental Material~\cite{Sl}. As tabulated in \textcolor{red}{Note S13} of the Supplemental Material~\cite{Sl}, 1H-WN$_2$ exhibits an exceptionally high room-temperature out-of-plane $\kappa_{\mathrm{L}}$ of 416~W\,m$^{-1}$\,K$^{-1}$, originating from it strong chemical bonds. This value approaches that of SiC calculated using the same method~\cite{PhysRevB.110.035205}, highlighting the excellent heat-dissipation capability of 1H-WN$_2$.

	In summary, we develop a hierarchical framework that combines high-throughput first-principles calculations with Boltzmann transport theory to identify high-mobility semiconductors. This approach identifies polar semiconductors $M$N$_2$ ($M$ = Mo and W) exhibit exceptionally high intrinsic room-temperature hole mobilities, reaching $10\,659$~cm$^2$\,V$^{-1}$\,s$^{-1}$ in 1H-WN$_2$. This exceptional transport performance originates from the concurrent suppression of polar-optical-phonon, acoustic-deformation-potential, and intervalley scattering. The remarkably weak polar-optical-phonon scattering follows from anomalously small Born effective charges and ultrahigh optical-phonon frequencies associated with covalent N--N bonds. The same robust N--N bonding produces a large bulk modulus, thereby mitigating acoustic-deformation-potential scattering. Moreover, strong spin--orbit coupling generates pronounced spin--valley locking at the valence-band maximum through symmetry-constrained hybridization between the W-$5d_{xy}$/$5d_{x^2-y^2}$ and N-$2p_x$/$2p_y$ orbitals, sharply restricting the available intervalley-scattering channels. These results establish 1H- and 2H-$M$N$_2$ as unexpectedly efficient hole conductors for high-performance electronics, challenge the conventional polarity--mobility tradeoff, and demonstrate crystal-structure engineering as a powerful route to controlling electron--phonon scattering.\\

	\textit{Acknowledgments}--Z.H., W.X., Z.X., and J.H. acknowledge the support of the National Science Foundation of China (Grant No. 12374024) and the Fundamental Research Funds for the Central Universities (No. FRF-BRA-26-007). R.T.P. and J.J.Z. acknowledge financial support from the National Natural Science Foundation of China (Grant Nos. 12574250 and 12104039), the Beijing Natural Science Foundation (Grant No. Z260002), and the National Key R\&D Program of China (Grant No. 2022YFA1403400).

	\bibliographystyle{aipnum4-1}
	\bibliography{ref}
	
\end{document}